\documentclass[conference]{IEEEtran}
\IEEEoverridecommandlockouts
\usepackage{cite}
\usepackage{amsmath,amssymb,amsfonts}
\usepackage{algorithmic}
\usepackage{graphicx}
\usepackage{textcomp}
\usepackage{xcolor}
\def\BibTeX{{\rm B\kern-.05em{\sc i\kern-.025em b}\kern-.08em
    T\kern-.1667em\lower.7ex\hbox{E}\kern-.125emX}}

\usepackage{comment}
\usepackage{soul}
\usepackage{subcaption}
\usepackage{graphicx}
\usepackage{caption}
\usepackage{url}
\usepackage{xcolor}
\usepackage{colortbl}
\usepackage{booktabs}
    
\begin{document}

\title{
Unraveling the Italian and English Telegram Conspiracy Spheres through Message Forwarding
}

\author{

\IEEEauthorblockN{Lorenzo Alvisi}
\IEEEauthorblockA{\textit{IMT School for Advanced Studies} \\
Lucca, Italy \\
lorenzo.alvisi@imtlucca.it}
\IEEEauthorblockA{\textit{Institute of Informatics and Telematics} \\
\textit{National Research Council (IIT-CNR)} \\
Pisa, Italy \\
lorenzo.alvisi@iit.cnr.it}
\and
\IEEEauthorblockN{Serena Tardelli}
\IEEEauthorblockA{\textit{Institute of Informatics and Telematics} \\
\textit{National Research Council (IIT-CNR)}\\
Pisa, Italy \\
serena.tardelli@iit.cnr.it \\
Corresponding author}
\and
\IEEEauthorblockN{Maurizio Tesconi}
\IEEEauthorblockA{\textit{Institute of Informatics and Telematics} \\
\textit{National Research Council (IIT-CNR)}\\
Pisa, Italy \\
maurizio.tesconi@iit.cnr.it}

} 

\maketitle

\begin{abstract}
Telegram has grown into a significant platform for news and information sharing, favored for its anonymity and minimal moderation. This openness, however, makes it vulnerable to misinformation and conspiracy theories. In this study, we explore the dynamics of conspiratorial narrative dissemination within Telegram, focusing on Italian and English landscapes. In particular, we leverage the mechanism of message forwarding within Telegram and collect two extensive datasets through snowball strategy. We adopt a network-based approach and build the Italian and English Telegram networks to reveal their respective communities. By employing topic modeling, we uncover distinct narratives and dynamics of misinformation spread. Results highlight differences between Italian and English conspiracy landscapes, with Italian discourse involving assorted conspiracy theories and alternative news sources intertwined with legitimate news sources, whereas English discourse is characterized by a more focused approach on specific narratives. Finally, we show that our methodology exhibits robustness across initial seed selections, suggesting broader applicability. This study contributes to understanding information and misinformation spread on Italian and English Telegram ecosystems through the mechanism of message forwarding. 
\end{abstract}



\section{Introduction}
Telegram has grown popular thanks to its commitment to anonymity, low moderation, and privacy, establishing itself as a significant hub for news and information.
Yet, the very features that attract users also open doors for misinformation to spread. 
In fact, Telegram's minimal content moderation serves as a double-edged sword. On the one hand, it fosters valuable information exchange on sensitive issues. On the other hand, this freedom creates fertile ground for the proliferation of conspiracy theories and misleading information to large audiences. 
For example, Telegram has emerged as hotspot for misinformation during critical political events, including elections in countries like Spain~\cite{tirado2023negative} Brazil~\cite{junior2021towards,cavalini2023politics}, and the United States~\cite{walther2021us}, challenging election integrity and promoting divisive ideologies. Similarly, the platform has served as a fertile environment for the spread of misinformation on topics such as the infodemic, pandemic, and other societal issues~\cite{ng2020analyzing,curley2022covid,tardelli2022cyber,zehring2023german}. Additionally, Telegram has been exploited by crypto investors to orchestrate large-scale market manipulations, including pump and dump schemes~\cite{xu2019anatomy,nizzoli2020charting}. 
The platform has also facilitated ideology radicalization~\cite{jost2023radical}, coordination of attacks, including those on Capitol Hill~\cite{hoseini2023globalization}, mobilizing protests~\cite{urman2021analyzing}, and the promotion of other conspiratorial narratives~\cite{schulze2022far,zehring2023german}, thus playing a crucial role in influencing public discourse and impacting democratic processes.
Discussing how these phenomena organize and characterize themselves is crucial for understanding the direction and evolution of public discourse and the factors influencing it. This understanding is vital not only for making online environments safer but also for grasping potential offline developments. This involves examining the dynamics within these platforms to identify how misinformation spreads, the community structures that support such narratives, and the implications for broader societal issues. This analysis can inform strategies to mitigate the spread of harmful content and foster a healthier public dialogue.
In this study, we focus on understanding the spread of conspiratorial narratives within Telegram communities through message forwarding, specifically within Italian and English language landscapes.
Message forwarding on Telegram involves sharing a message from one chat directly into another, serving as a critical mechanism for distributing content across different user groups. We hypothesize that forwarded messages not only distribute content but also signal homophily, that is shared interests and beliefs, among community members, similar to how the diffusion of invite links has been studied in the past~\cite{anderson2015global,nizzoli2020charting}.

\textbf{Contributions.}
We first collect data from Telegram by leveraging message forwarding. Starting from selected initial chats as seeds, we perform iterative, snowball sampling and expand the data by iteratively identifying and retrieving new chats and messages. 
We collect two large datasets: the Italian dataset covers the period from January 1, 2024, to February 13, 2024, and comprises more than 1K chats and 3.4M messages. Meanwhile, the English dataset spans from January 1, 2024, to February 20, 2024, and consists of more than 600 chats and 5M messages.
We build two Telegram networks based on message forwarding, identify key communities and employ topic modeling to characterize their discussions and understand the specific narratives.
We show that the Italian landscape of conspiracy theories forms a network involving religious groups, Russian influences, anti-vaccination proponents, and news source of varying reliability. In contrast, the English landscape appears more tied to structured conspiracies, involving ties with cryptocurrency scams.  
Finally, we validate our method by showing that our findings does not depend on the initial seeds, offering a new lens through which to examine the flow of information and misinformation. 

We summarize our main contributions in the following: 
\begin{itemize}
    \item We leverage message forwarding to collect two extensive Telegram conspiracy-related datasets, including channels, groups -- often overlooked in existing literature, and messages. For the first time, we also incorporate linked chats, which are two-tiered structures consisting of channels linked to their respective groups.
    \item We characterize conspiratorial narratives within Telegram communities, focusing on both English and Italian spheres, shedding light on Italian Telegram dynamics not extensively explored in existing literature.
    \item We highlight differences in conspiracy theory landscapes between Italian and English-speaking communities, revealing the presence of diverse news sources playing varied roles in shaping discourse, and exploring the connections among various conspiracy theories within these groups.
    \item We show that forwarded messages serve for content distribution and signal community homophily, and that our insights do not overly depend on the initial selection of seeds, suggesting the robustness and broad applicability of our methodology.
\end{itemize}

\section{Related Works}

\subsection{Overview of Telegram data collection methods} 

Several studies relied on message forwarding to collect data from Telegram. For example, the authors in~\cite{lamorgia2023tgdataset} aimed to create the largest collection of English Telegram channels, spanning a wide range of diverse topics, with their analysis primarily centered on dataset statistics. In contrast, research in~\cite{willaert2023computational} analyzed communities by building user networks from forwarded messages, and exploring the narratives within. Similarly, research in~\cite{BovetGrindrod2022ukfarright} and~\cite{baumgartner2020pushshift} followed a snowball characterized specific English-speaking Telegram communities of channels. Our study, however, expands on this foundation by incorporating not just channels but also groups into our analysis. Specifically, we uniquely consider the \textit{linked chat} feature on Telegram, where a channel is directly connected to a group. To the best of our knowledge, this is the first research effort to include this duality feature in literature.
Other studies adopted snowball approaches on Telegram, focusing on different elements like mentions~\cite{telegramforwardBT} or invite links -- special URLs that allow users to join channels~\cite{glenski2019characterizing,nizzoli2020charting}. For example, the research in~\cite{nizzoli2020charting} analyzed how fraudsters used these invite links in scam channels to attract large audiences, highlighting the significance of invite link diffusion patterns for identifying homophily and shared interests within online communities. In a similar way, the study in~\cite{BovetGrindrod2022ukfarright} explored the concept within far-right communities, proposing that Telegram groups act as echo chambers and that the sharing of forward links suggests a level of homophily. Building on this, our research seeks to further explore the utility of forward links in content distribution and their ability to reveal homophily among users.
Lastly, other studies employed different data collection strategies, such as gathering messages from an initial set of seeds without employing a snowballing approach~\cite{xu2019anatomy,Avalle2024}. These studies primarily aim to illustrate the unfolding of specific events, like instances of toxicity or fraud schemes.

\subsection{Studies of conspiracy in Italian and English Telegram discussions} 

Conspiracy theories have been identified and analyzed across various platforms, thriving in numerous online environments~\cite{calamusa2020twitter,kim2021propagation,engel2022characterizing,gambini2024anatomy}, including Telegram.
The majority of the research on Telegram has focused on conspiracy theories within English-speaking discussions, including studies on the pandemic~\cite{vergani2022hate}, the far-right~\cite{davey2021inspiration,schulze2022far}, and the QAnon movements~\cite{hoseini2023globalization}. Notably, the QAnon conspiracy, in particular, has been linked to a wide range of conspiratorial narratives, highlighting its broad influence~\cite{willaert2023computational,greer2024belief,tuters2022deep}. Building upon this works, our study extends the examination of conspiracy discourse in English-speaking communities, especially QAnon and its current connections with other narratives.
On the other hand, the realm of conspiracy theories within Italian-speaking Telegram communities remains largely unexplored. The Italian conspiracy ecosystem on Telegram came to the spotlight during the COVID-19 pandemic~\cite{vergani2022hate}, as protest movements gained significant social momentum, leading to widespread protests~\cite{spitale2022concerns}, movements having ties with Italian alt-right, a phenomenon observed also in other European countries~\cite{zehring2023german}. Other studies focused into the Italian QAnon disinformation infrastructure~\cite{pasquetto2022disinformation}, highlighting the closed nature of these communities within the Italian sphere, similarly to English-speaking environments~\cite{willaert2023computational}. Despite these insights, a comprehensive understanding of the broader conspiracy landscape in Italy remains unexplored. Our study seeks to fill this gap by examining the connections between various conspiracy narratives in Italian-speaking Telegram communities, and comparing them with English-speaking communities.



\section{Methodology}

\subsection{Designing and collecting the dataset}

\begin{figure}[!t]
\centering
    \begin{minipage}{.24\textwidth}%
        \centering
        \includegraphics[width=\textwidth]{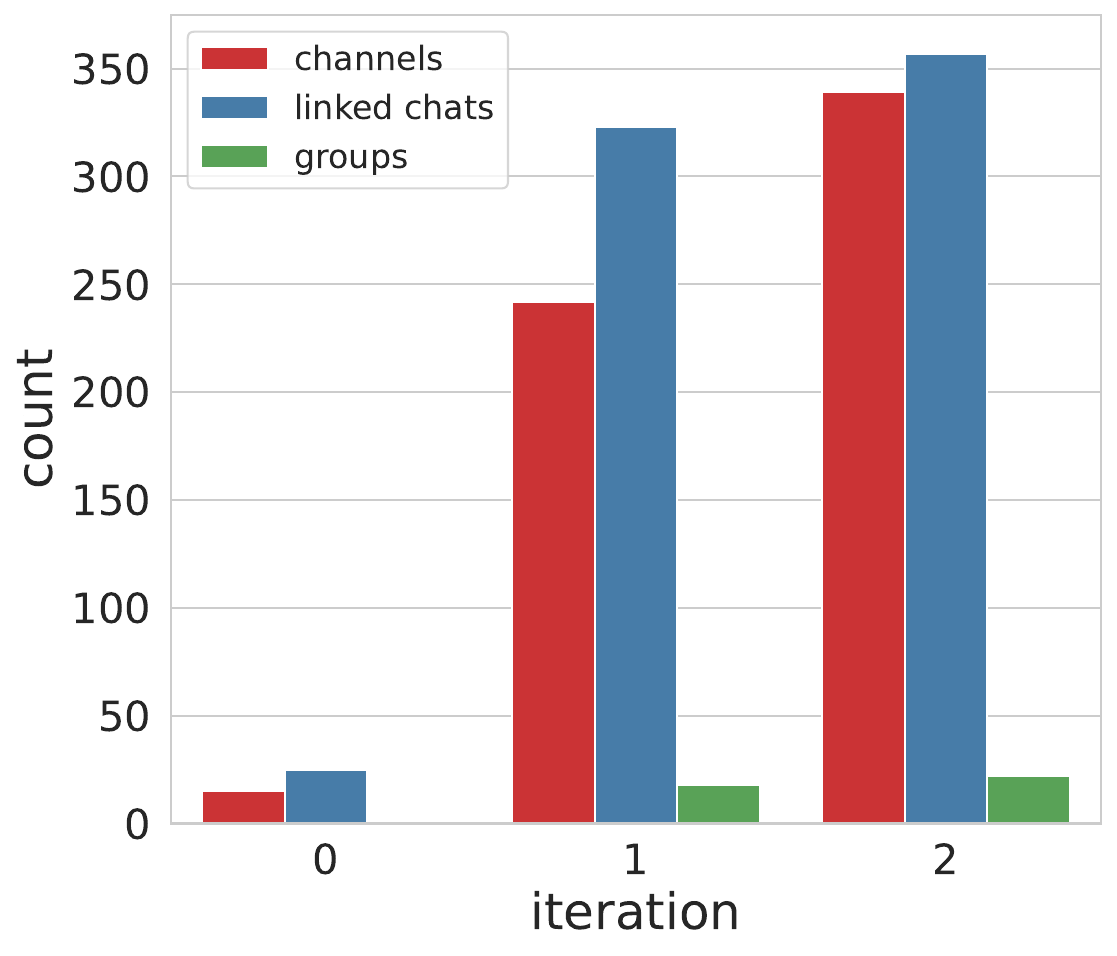}
        \subcaption{IT}
    \end{minipage}%
    \begin{minipage}{.24\textwidth}%
        \centering
        \includegraphics[width=\textwidth]{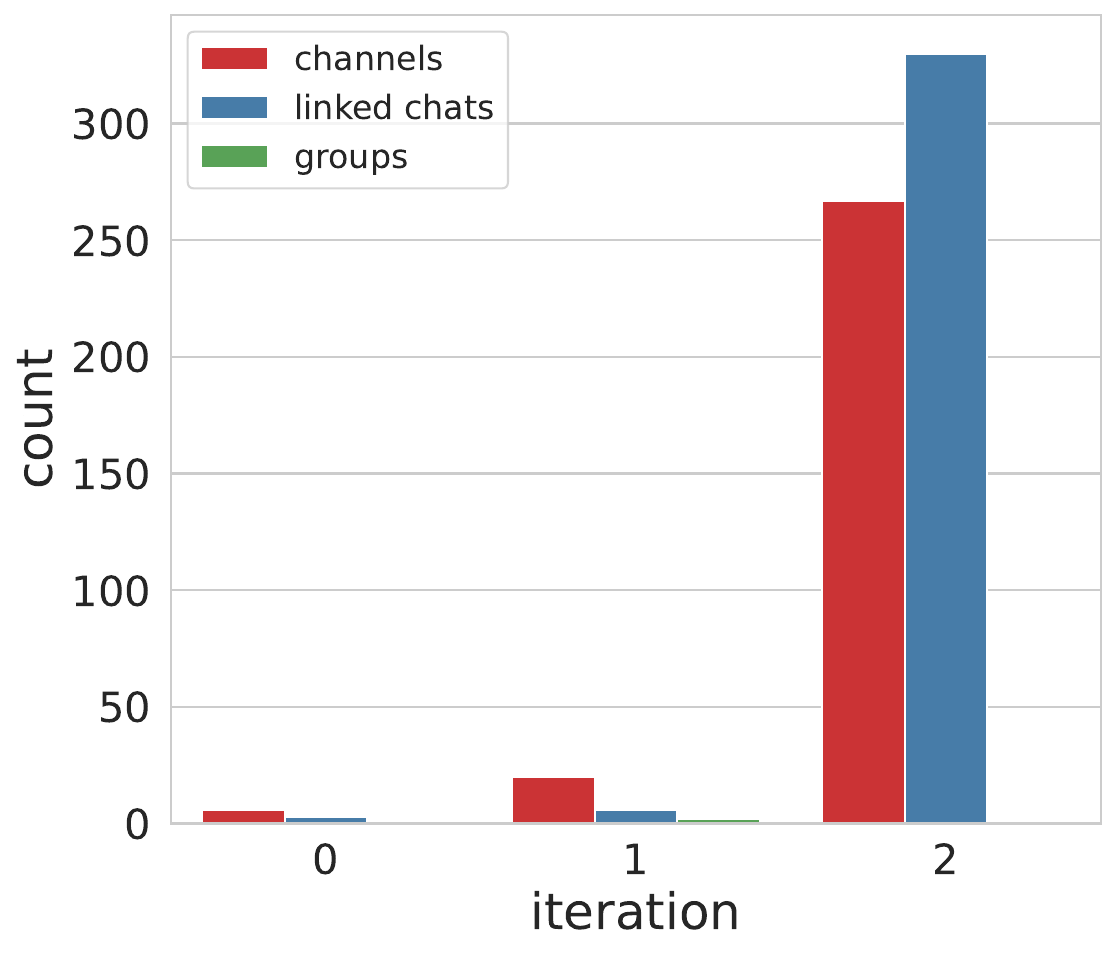}
        \subcaption{EN}
    \end{minipage}%

\caption{Retrieved chats by iteration}
    \label{iteration process en}
\end{figure}

\paragraph{Telegram terminology}\label{preliminaries}

Telegram offers a variety of chat types. \textit{Channels} are unidirectional chats where typically only administrators broadcast content to an audience that cannot interact directly. \textit{Groups} are chat rooms where all members have permission to share contents by default and interact with each other. \textit{Supergroups} are a variation of groups, differentiated mainly by administrative powers and member limits. However, for our study, we treat them as equivalent to regular groups since these differences are not relevant to our analysis.
A notable feature in Telegram is the ability for channel admins to link a channel to a corresponding group, creating a two-tiered structure known as \textit{linked chat}. 
In this structure, a channel enables any user, whether a follower or not, to reply directly to each post. Simultaneously, the associated group houses these conversational threads and operates as a standard group. This composite structure allows unrestricted interaction on the channel's posts and fosters broader discussion within the group.
For the scope of our paper, we consider public channels, groups, and linked chats. We use the term \textit{chat} interchangeably to refer to all three types.
As mentioned, we highlight a key Telegram feature, that is the ability for users to share posts and messages from one chat to another via \textit{message forwarding}. This feature preserves the original chat’s information, effectively creating a bridge between chats and facilitating the discovery and retrieval of connected content.

\paragraph{Data collection approach}
We retrieve two distinct Telegram datasets pertaining to conspiracy discussions in Italian and English using the following approach.
We employ a snowball technique focused on message forwarding, a method previously used in several papers for channel retrieval~\cite{telegramforwardBT,lamorgia2023tgdataset}. For the first time, we expand this technique to include groups and linked chats. We begin by selecting seed chats known for conspiracy content. For the Italian discussions, we select seeds through keyword on tgstats.com, a platform that provides a categorized catalog of existing Telegram chats. We focus on terms associated with pandemic conspiracy theories, identifying 43 Italian chats related to conspiracies as seeds. Similarly, for the English seeds, we use tgstats and search for keywords associated with the QAnon conspiracy, resulting in 20 seed chats.
We start from two different conspiracy theories to anchor our study in the specific cultural and linguistic contexts, ensuring a focus on the conspiracy sphere and exploring how these conspiracies expand and evolve in these settings.
We leverage Telegram APIs to collect messages. Starting with seed chats at iteration~0, we parse messages to identify forwarded messages, following them to retrieve new chats and their messages in subsequent iterations. We only add new chats that meet our language criteria, either Italian or English, determined by the most frequently detected language in their messages. Our data collection concludes after iteration~2.

\paragraph{Datasets overview}

\begin{figure}[!t]
\centering
    \begin{minipage}{.24\textwidth}%
        \centering
        \includegraphics[width=\textwidth]{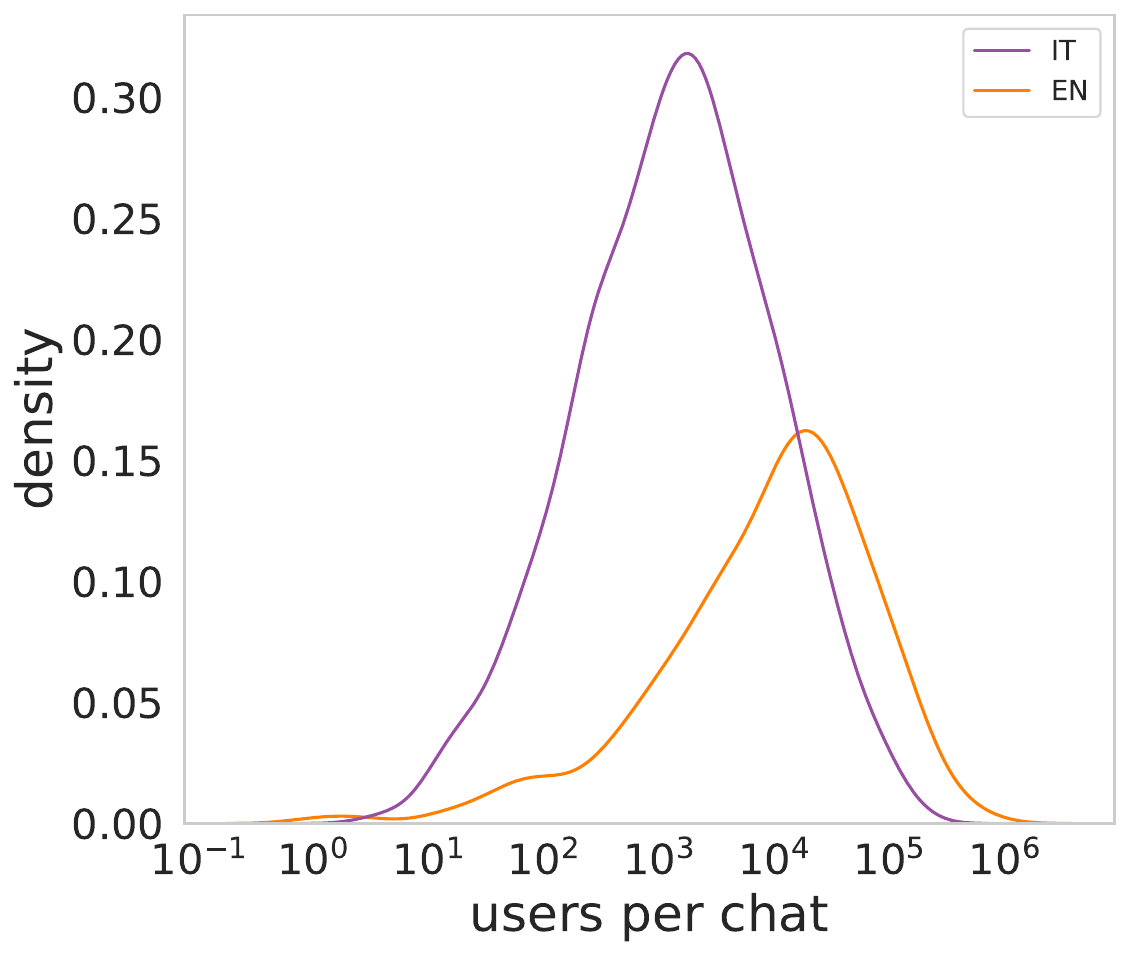}
        \subcaption{Users}
    \end{minipage}%
    \begin{minipage}{.24\textwidth}%
        \centering
        \includegraphics[width=\textwidth]{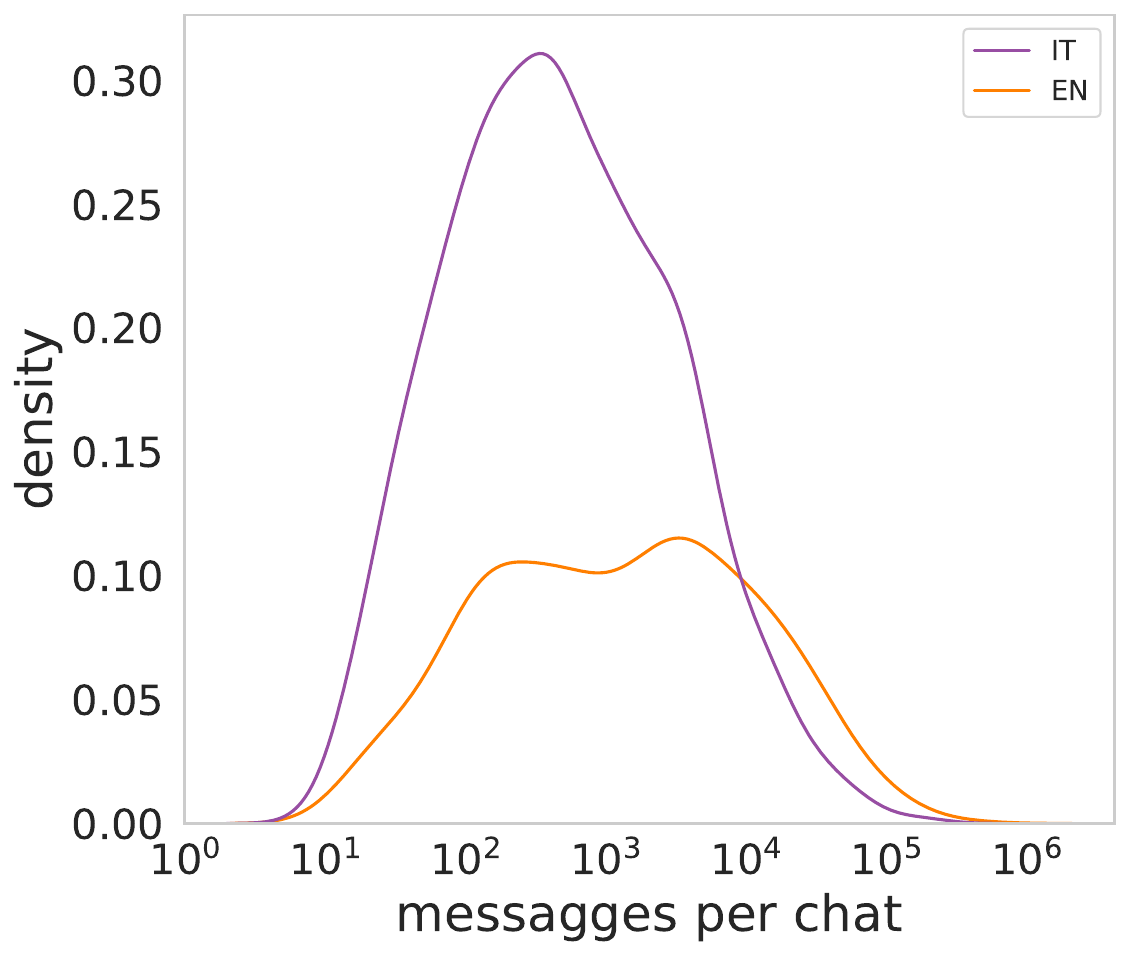}
        \subcaption{Messages}
    \end{minipage}%
    
    \caption{Distribution of users and messages per chat}
    \label{user-msg-distrib}
\end{figure}

Using the aforementioned approach, we collect two large datasets: the Italian dataset, covering the period from January 1, 2024, to February 13, 2024, includes $1,346$ chats, containing a total of 3.4M messages. Meanwhile, the English dataset, spanning from January 1, 2024, to February 20, 2024, comprises $634$ chats, including a total of 5M messages. Figure~\ref{iteration process en} shows the number of chats per type collected at each iteration of our snowball crawling strategy. Predominantly, linked chats are more prevalent at each stage, while standalone groups are less used in these contexts. 
We analyze the distribution for both the number of users and the number of comments per chat. Linked chats required a specialized approach for analysis. For message counts, we aggregate the total number of messages across both linked chats. For user counts, we consider the higher number of subscribers, whether from the channel or its linked group.
As shown in Figure~\ref{user-msg-distrib}, we observe that the log-number of users and messages within the chats exhibit a gaussian distribution, contrasting with the typical heavy-tailed distribution of conversational trees documented in prior research~\cite{Avalle2024}. This variation could imply that linked chats and groups, being more similar to chat rooms than traditional social media feeds, might exhibit different behaviors. Alternatively, it could suggest that our snowballing techniques could miss smaller chats, thus filtering out less influential ones.

\subsection{Building the networks/uncovering communities}
The message forwarding mechanism enables us to construct a directed weighted graph $\mathcal{G}=(\mathcal{N},\mathcal{E})$, where $\mathcal{N}$ represents the set of nodes and $\mathcal{E}$ the set of edges. In this graph, nodes correspond to chats, which include unlinked channels, unlinked groups, and linked chats. For any two nodes $u, v \in \mathcal{N}$, the weight of the edge $w_{e_{u,v}} \in \mathcal{E}$ is determined by the number of messages forwarded from chat $u$ to chat $v$. To prevent loops, forwards from a chat to itself, including within linked chats, are excluded. This exclusion is crucial as, in linked chats, each message from the channel is automatically forwarded to the associated group to form conversational trees.
The Italian network consists of $1,346$ nodes and $35,802$ edges, and the English network comprises $634$ nodes and $24,546$ edges. We employed community detection within our graph using the Louvain algorithm tailored for directed graphs~\cite{directedlouvain}, focusing only on communities with more than 10 chats to ensure the robustness of our findings.
\section{Results}
The application of our methodology brought to the detection of multiple communities within the Italian and English Telegram conspiracy landscapes.
In the following we shed light on the activity and dissemination patterns of the communities.

\begin{table*}[t]
	\footnotesize
	\centering
	\begin{tabular}{lrcl}
	   
		\toprule
		\textbf{Italian communities} & \textbf{Size} && \multicolumn{1}{c}{\textbf{Top words in ranked topics}} \\
            \midrule
            \texttt{Freedom}	  & \scriptsize{\texttt{297}}   && \textit{liberademocrazia, dissenso, geopolitica, democrazia, anonimato, governo, imporre, controllare} \\ 
	\rowcolor{gray!10} 
            \texttt{Warfare}	& \scriptsize{\texttt{261}} && \textit{ucraino, yemen, internazionale, biden, geopolitica, gaza, russia}  \\
            \texttt{ConspiracyMix}	 & \scriptsize{\texttt{249}}    &&  \textit{governo, pandemia, salute, genocidio, storia, agricoltore, protesta, biden, trump} \\ 
        \rowcolor{gray!10} 
            \texttt{ConspiracyMix2}	 & \scriptsize{\texttt{188}}        && \textit{trump, epstein, agricoltore, alimentare, guerra, vaccino, covid, sicurezza, dissenso, diritto} \\ 
            \texttt{NewsSource}	 & \scriptsize{\texttt{102}}    && \textit{warrealtime, internazionale, informazione, media, ministero, affermare}\\
        \rowcolor{gray!10} 
            \texttt{Politics}	 & \scriptsize{\texttt{73}}    &&  \textit{politica, economico, governo, presidente, italia, europeo, ministro, pubblico, carabiniere} \\
            \texttt{AltNews}	  & \scriptsize{\texttt{52}}   &&  \textit{bankers, informazione, censura, globalista, società, imporre, morte, libertà, controllare}\\
        \rowcolor{gray!10} 
            \texttt{Fight}	 & \scriptsize{\texttt{43}}    &&  \textit{popolare, lotta, civile, collegare, verità, libertà, importante, agire} \\
            \texttt{Novax}	  & \scriptsize{\texttt{37}}   &&  \textit{dissenso, vaccinazione, bambino, studio, mortalità, controinformazione, salute, rischiare}\\
        \rowcolor{gray!10} 
            \texttt{Religious}	 & \scriptsize{\texttt{14}}    &&  \textit{gesù, valore, sacramento, pentire, rinascere, invidia, esorcista, miracolosamente, guarigione} \\
            \texttt{Spiritual}	  & \scriptsize{\texttt{12}}   &&  \textit{awakening, riflessione, luce, conscience, inspirations, meditation} \\
            
            \midrule
            \textbf{English communities} & &&  \\
            \midrule
        \rowcolor{gray!10}
            \texttt{QAnonCrypto}	 & \scriptsize{\texttt{119}}    && \textit{trump, god, control, chadgptcoin, coin, btc, pump, dump, money, official} \\
            \texttt{Warfare}	 & \scriptsize{\texttt{117}}    && \textit{ukrainian, attack, military, israel, defense, missile, rhetoric, soldier} \\
        \rowcolor{gray!10}
            \texttt{QAnonHealth}	  & \scriptsize{\texttt{103}}   && \textit{trump, god, child, food, cancer, parasite, health, weapon, medical, water} \\
		\texttt{CHScams}	& \scriptsize{\texttt{89}}     && \textit{transfer, money, deposit, click, payment, win-win, card, finance} \\
        \rowcolor{gray!10}
            \texttt{QAnon}	  & \scriptsize{\texttt{73}}   && \textit{endhumantrafficking, minor, abuse, police, evil, control, trump, god} \\
            \texttt{ConspiracyMix}	   & \scriptsize{\texttt{52}}  && \textit{kaplan, dogedesigner, elon, war, trump, heaven, biden, border, ballot, court} \\ 
        \rowcolor{gray!10}
            \texttt{Covid}	 & \scriptsize{\texttt{30}}    && \textit{vaccine, covid, health, body, food, cancer, doctor, government}   \\
            \texttt{OldSchoolConsp}	 & \scriptsize{\texttt{22}}    && \textit{weird, shit, ufo, alien, paranormal, time, experience, consciousness} \\
		\bottomrule
	\end{tabular}
	\caption{Topics identified by Corex models. For each community, The words listed in each row correspond to the top-ranked terms associated with that community, as determined by the Corex algorithm. This highlights the main terms and topics prevalent within each community.} 
	\label{tab:corex-communities}
\end{table*}


\subsection{Uncovering narratives}
Here, we present summary information for each community, alongside their main narratives. We uncover the main topics of discussion within each community through a comprehensive analysis approach. This involves utilizing topic modeling techniques, channel information, and examining TF-IDF weighted hashtags used by each community. By leveraging these diverse methods, we aim to offer valuable insights into the unique themes and narratives that shape the discourse within each community.
To perform topic modeling, we adopted a state-of-the-art algorithm known as Anchored Correlation Explanation (CorEx)~\cite{gallagher2017anchored}.
Unlike traditional methods like Latent Dirichlet Allocation (LDA), CorEx identifies hidden topics within a collection of documents without assuming any particular data generating model. Instead, it leverages the dependencies of words in documents through latent topics, by maximizing the total correlation between groups of words and the respective topic, ensuring greater flexibility~\cite{gallagher2017anchored}. 
We applied unsupervised CorEx, in order to discover topics spontaneously emerging from our data. Given that our network consists of chat platforms, with each chat having a one-month history, we trained separate models for each community. We utilized the chat messages as corpora to capture the full spectrum of topics discussed within each community. This approach allows us to comprehensively explore the range of topics present in each community's discourse.
After experimenting with different configurations, we set the expected number of topics to 10, since additional topics were adding negligible correlation to the learned models. Finally, we ranked the obtained topics according to the fraction of the total correlation that they explain. Results are presented in Table~\ref{tab:corex-communities} and discussed as follows. 

\textbf{Italian Narratives.} The Italian-speaking communities are presented as follows and presented in Table~\ref{tab:corex-communities}, ordered by decreasing number of members: 
\begin{itemize}
    \item \texttt{Freedom}: This community is centered around concepts of liberal democracy and dissent, discussing geopolitical topics, democracy, anonymity in governance, and control-related issues.
    \item \texttt{Warfare}: A community concerned with international warfare, particularly focusing on the Ukrainian conflict and Russian propaganda.
    \item \texttt{ConspiracyMix}: A community that discusses various conspiracy theories involving government actions, health-related topics such as the pandemic, and foreign political figures.
    \item \texttt{ConspiracyMix2}: Similar to ConspiracyMix, this community spans across  conspiracy theories, touching on warfare, vaccines, COVID-19, farmers' protests, and QAnon.
    \item \texttt{NewsSource}: A community that encompasses a spectrum of information sources ranging from conspiracy theory-driven outlets to reputable journalistic sources (e.g., ``IlSole24Ore,'' ``IlFattoQuotidiano''). This convergence reflects the dynamics of conspiratorial contexts, where genuine information is often filtered through a conspiratorial lens, shared, and discussed alongside news from international sources, with an emphasis on media scrutiny and critique~\cite{uscinski2018study,mahl2022conspiracy}.
    \item \texttt{Politics}: A political community discussing economic issues, government policies and European affairs.
    \item \texttt{AltNews}: A community focused on counter-information and alternative news sources, focusing on issues of censorship, globalism, and societal control.
    \item \texttt{Fight}: A community engaged in civil struggles, emphasizing the importance of truth, freedom, and action in the face of societal challenges.
    \item \texttt{Novax}: A community characterized by dissent against vaccinations, health studies, health risks, and mortality rates.
    \item \texttt{Religious}: A community centered on Italian religious values, discussing Jesus, sacraments, and other themes of rebirth, envy, exorcism, and miraculous healing.
    \item \texttt{Spiritual}: A community centered on spiritual topics, such as spiritual awakening and meditation.
\end{itemize}

These communities all circle around conspiracy theories, each one with its own angle, with alternative information challenging mainstream narratives to news source offering more traditional views.
In addition, conspiracy narrative ties to religiosity, alternative health, and conspiratorial thinking, as observed in literature for English-speaking groups~\cite{willaert2023computational,greer2024belief,tuters2022deep}.
Exploring these groups gives us insight into the Italian conspiracy ecosystem on Telegram, a subject that is relatively unexplored in existing literature.

\textbf{English Narratives.} While our focus thus far has centered on Italian-speaking communities, here we present the English ones. Examining English-speaking communities allows us to provide valuable comparative insights into conspiracy theories in different cultural contexts. 
The English-speaking communities are presented as follows:
\begin{itemize}
    \item \texttt{QAnonCrypto}: A community where conspiracy discussions are hijacked by the cryptocurrency world, featuring themes of various coins and fraudulent schemes like pump and dump~\cite{xu2019anatomy}. Indeed, prior research has explored the involvement of cryptocurrency in discussions, noting the frequent presence of cryptocurrency and finance-related tags within QAnon-related themes~\cite{dilley2022qanon}. In fact, belief in conspiracy theories plays a role in people's decisions to invest in cryptocurrency, as people exhibiting cunning traits and a distrustful stance toward government are more likely to favor cryptocurrency as an investment option~\cite{martin2022dark}.
    \item \texttt{Warfare}: A community similar to its Italian counterpart, focusing on the Ukrainian conflict, military issues, and other war rhetoric.
    \item \texttt{QAnonHealth}: A community where QAnon conspiracy theories intersect with health concerns, discussing food, cancer, and parasites, along with other medical aspects.
    \item \texttt{CHScams}: A community that relies on conspiracy theory discussions to promote financial scams and fraudulent activities in Chinese language. The terms listed in the table are translated from Chinese to English.
    \item \texttt{QAnon}: This community focuses on pure QAnon conspiracy theories, involving topics such as child abuse, government control, and political figures.
    \item \texttt{ConspiracyMix}: This community discusses various conspiracy theories, with a focus on legal issues as seen in terms like ``court,'' while also touching the cryptocurrency sphere (e.g., ``DodgeCoin,'' ``Elon''). Discussions also involve Judge Lewis Kaplan, who presided over both Trump's federal defamation trial\footnote{\url{https://www.nytimes.com/2023/04/27/nyregion/who-is-lewis-kaplan-judge-in-carroll-case-against-trump.html}} and Sam Bankman-Fried's cryptocurrency fraud trial\footnote{\url{https://www.bloomberg.com/news/articles/2022-12-27/bankman-fried-case-reassigned-to-us-judge-lewis-kaplan-in-ny}}.
    \item \texttt{Covid}: A community centered around discussions of COVID-19, vaccine skepticism, and related health and governmental issues.
    \item \texttt{OldSchoolConsp}: A community focused on traditional conspiracy topics such as UFOs, aliens, the paranormal, and discussions of time and consciousness.
\end{itemize}

\begin{figure*}[!t]
    \centering
    \begin{minipage}[b]{0.35\textwidth}
        \includegraphics[width=\textwidth]{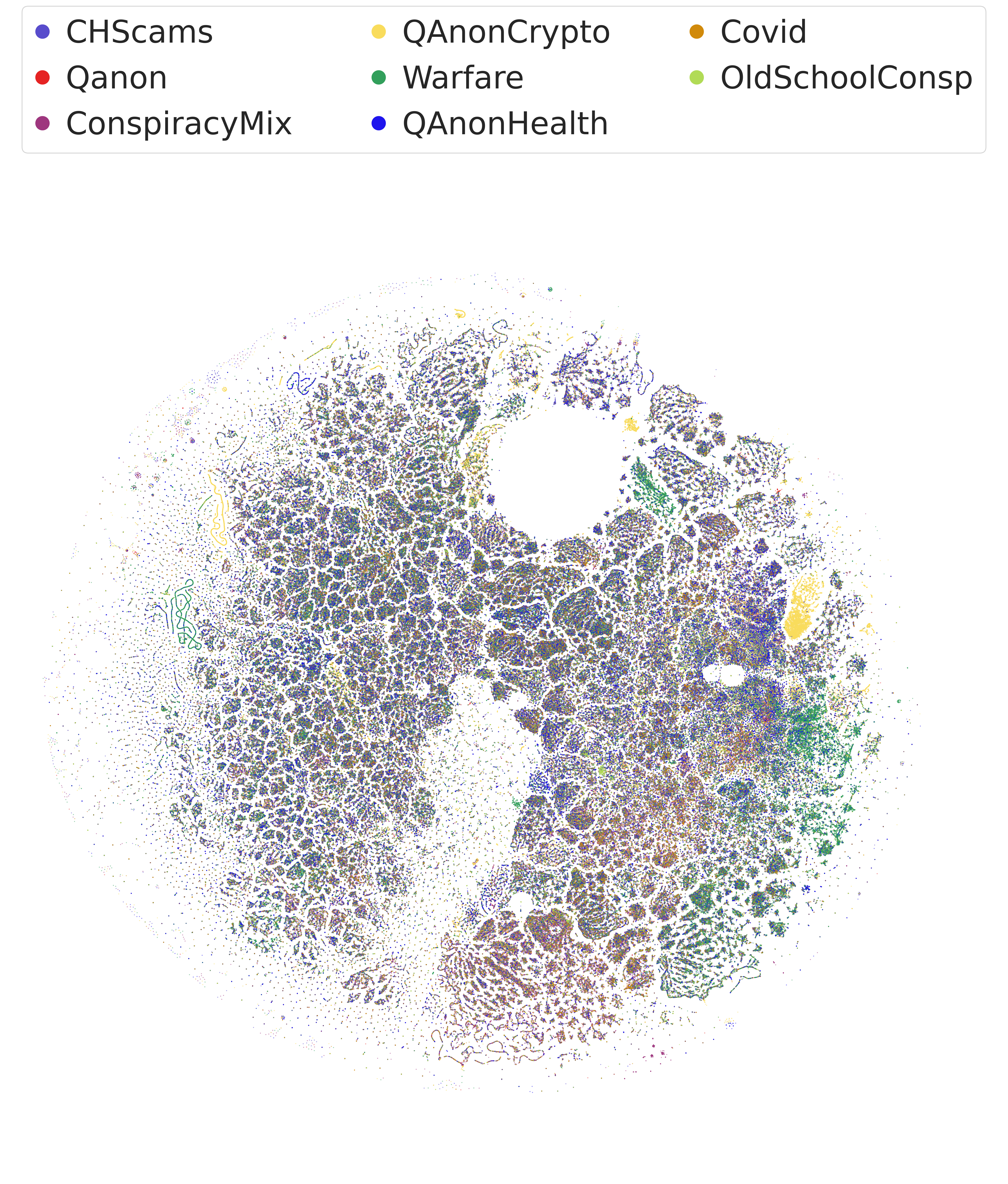}
        \caption{t-SNE representation of message distribution by topic in the EN Dataset}\label{fig:tsne}
        
    \end{minipage}
    \hfill
    \begin{minipage}[b]{0.64\textwidth}
        \vspace{5mm} 
        \begin{subfigure}{0.24\textwidth}
            \includegraphics[width=\textwidth]{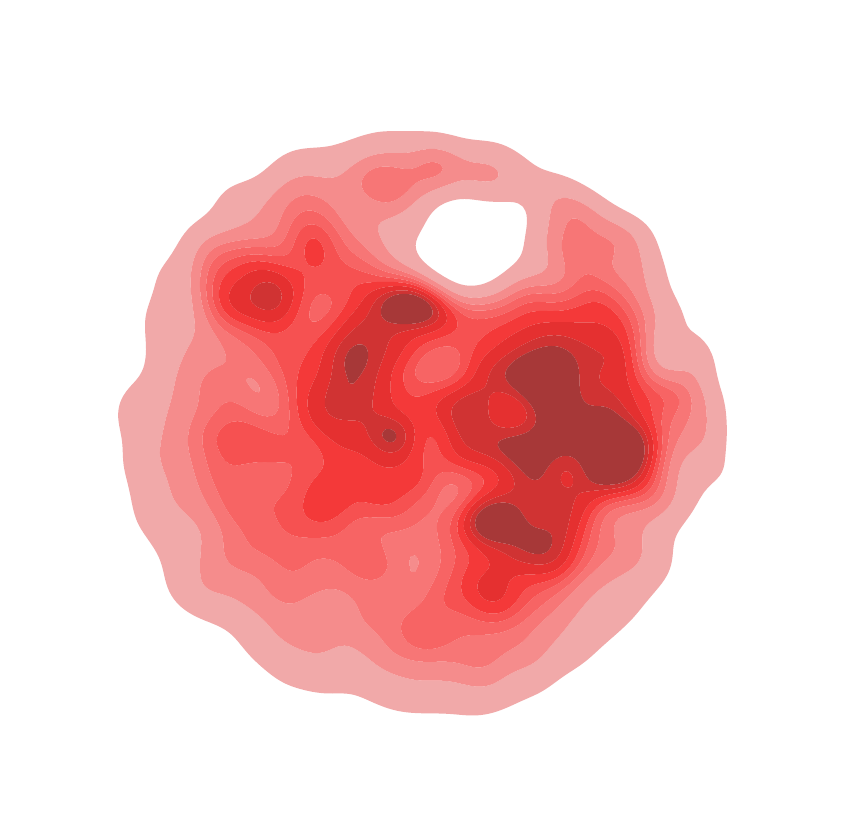}
            \subcaption{\texttt{QAnon}}
        \end{subfigure}
        \begin{subfigure}{0.24\textwidth}
            \includegraphics[width=\textwidth]{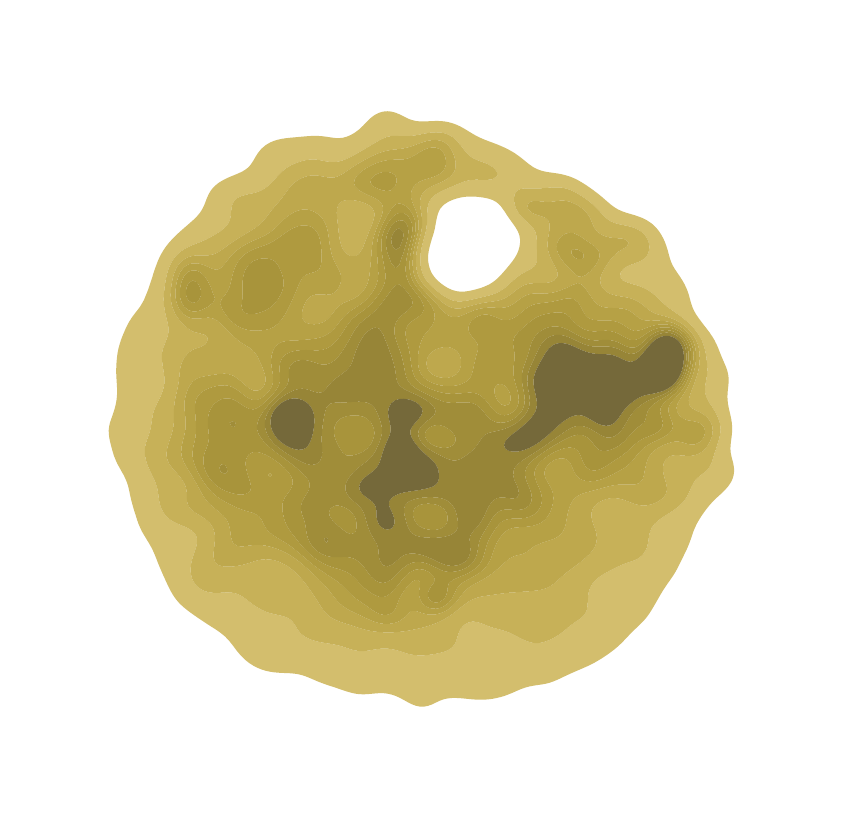}
            \subcaption{\texttt{QAnonCrypto}}
        \end{subfigure}
        \begin{subfigure}{0.24\textwidth}
            \includegraphics[width=\textwidth]{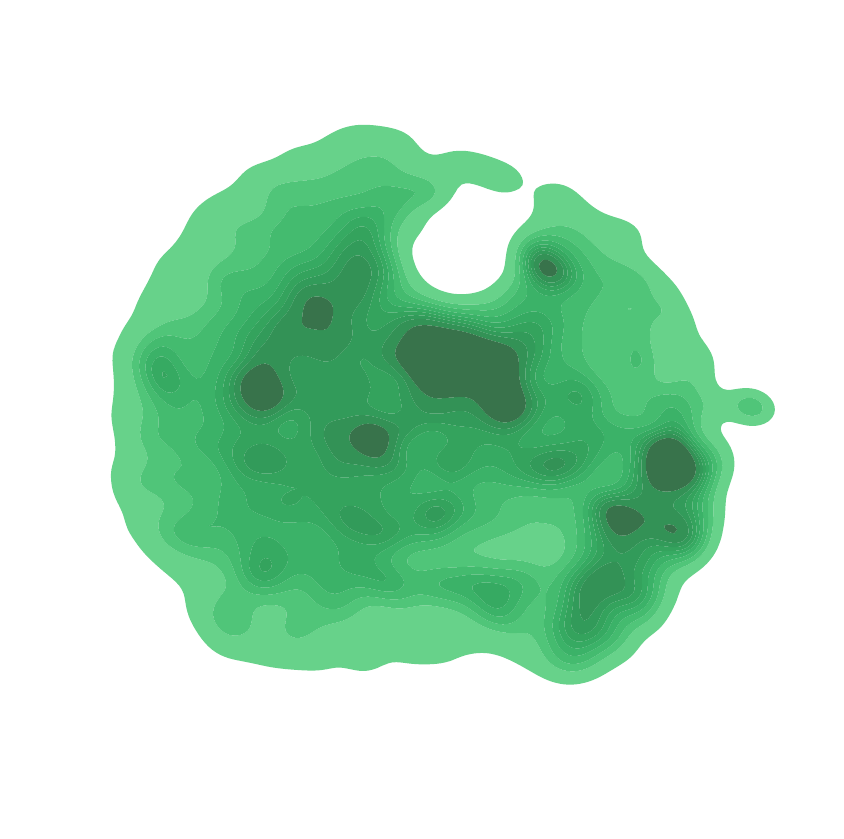}
            \subcaption{\texttt{Warfare}}
        \end{subfigure}
        \begin{subfigure}{0.24\textwidth}
            \includegraphics[width=\textwidth]{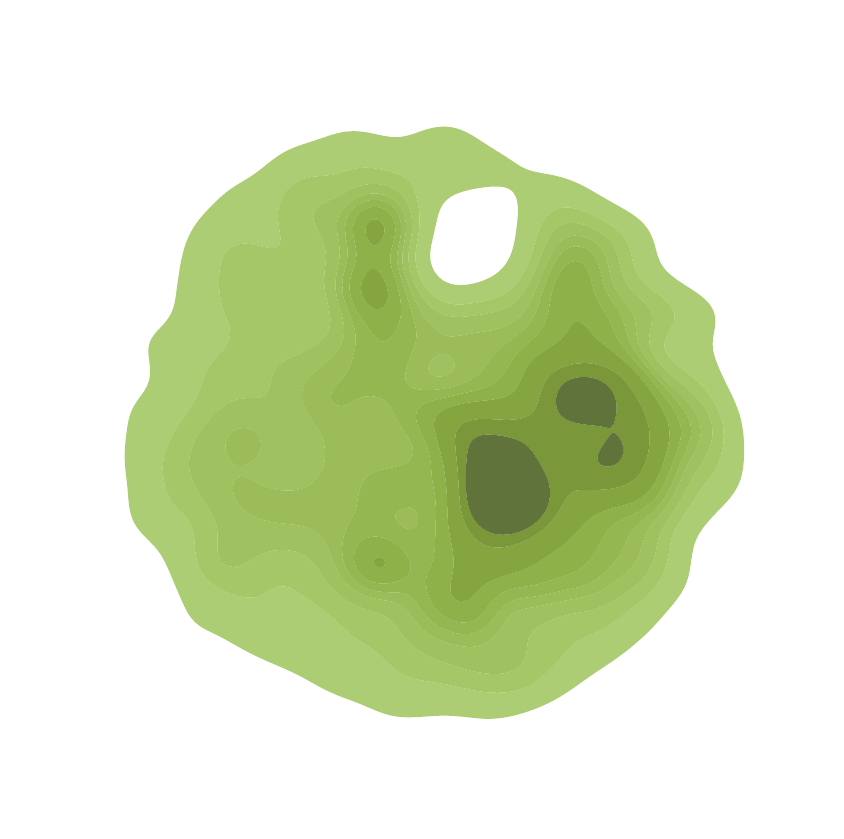}
            \subcaption{\texttt{OldSchool}}
        \end{subfigure}
        
        \caption{KDE of message topics for different EN communities}\label{fig:kde-en}
        
        \vspace{5mm}      
        \begin{subfigure}{0.21\textwidth}
            \includegraphics[width=\textwidth]{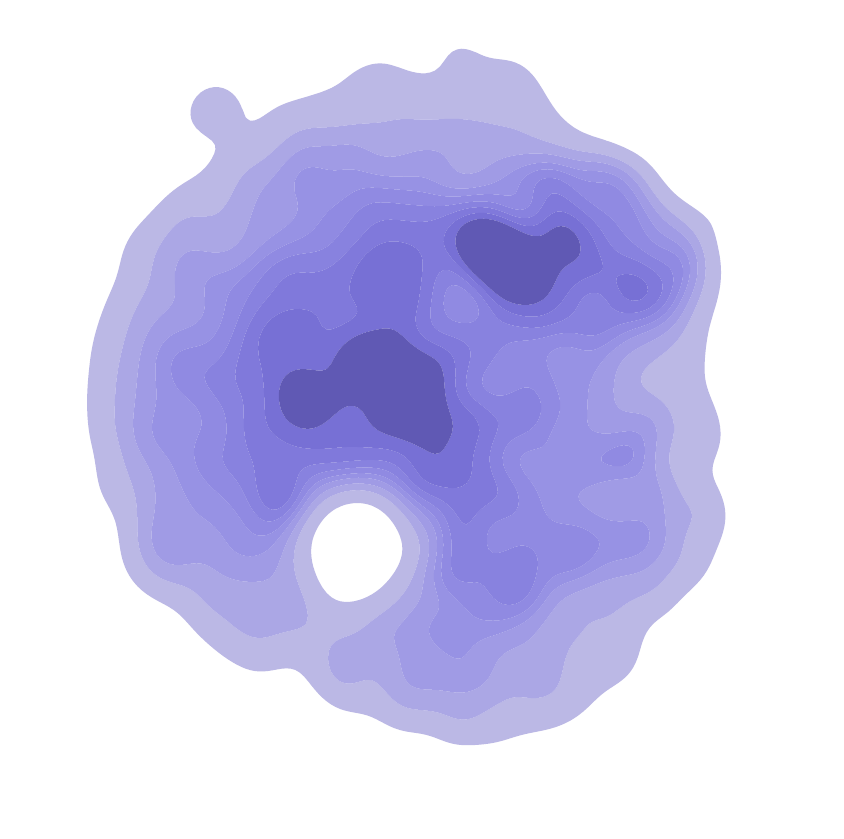}
            \subcaption{\texttt{NewsSource}}
        \end{subfigure} \hfill
        \begin{subfigure}{0.21\textwidth}
            \includegraphics[width=\textwidth]{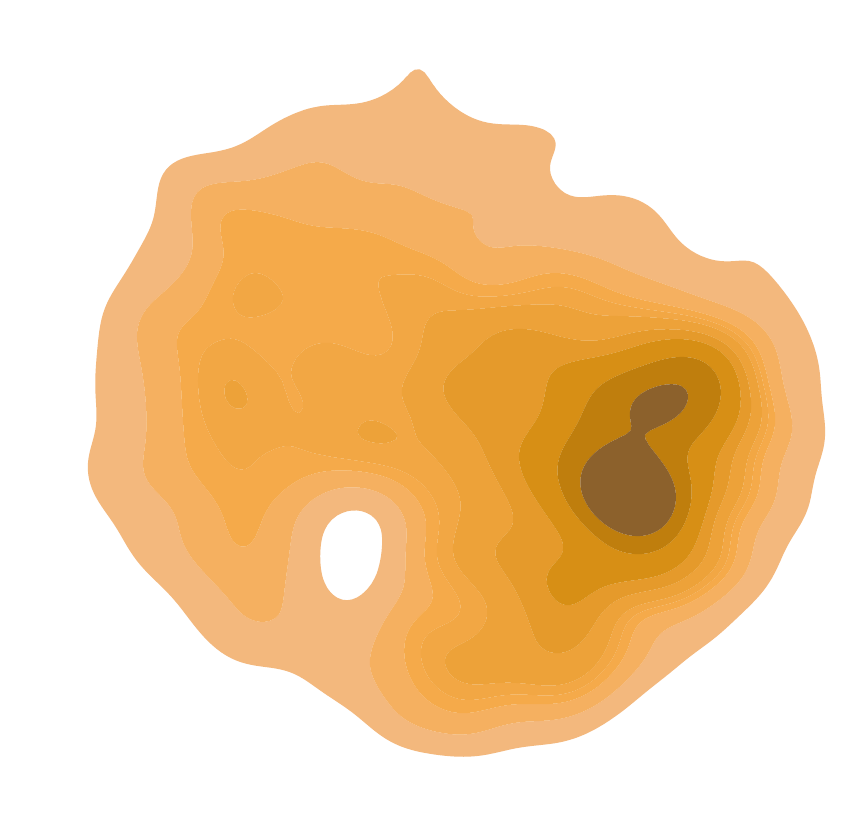}
            \subcaption{\texttt{AltNews}}
        \end{subfigure} \hfill
        \begin{subfigure}{0.21\textwidth}
            \includegraphics[width=\textwidth]{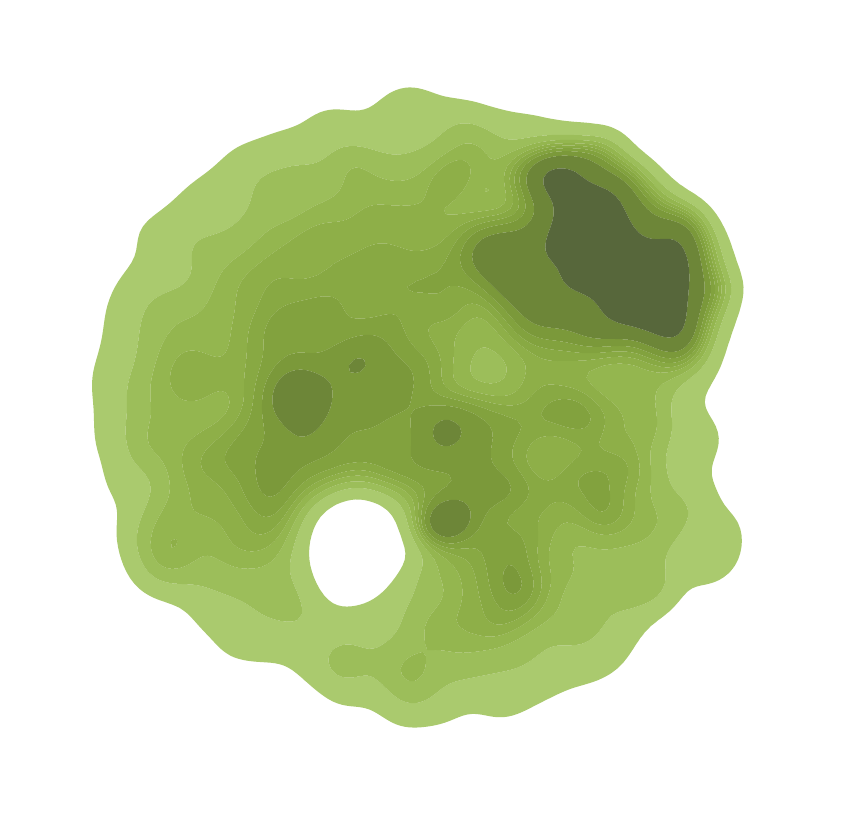}
            \subcaption{\texttt{Warfare}}
        \end{subfigure} \hfill
        \begin{subfigure}{0.21\textwidth}
            \includegraphics[width=\textwidth]{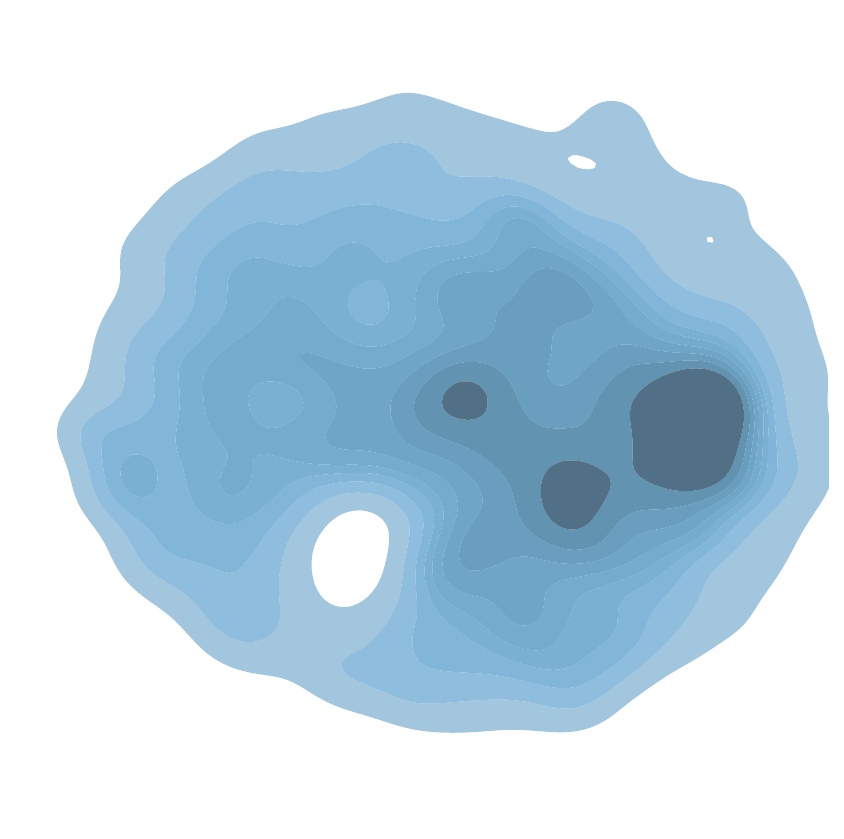}
            \subcaption{\texttt{Novax}}
        \end{subfigure}\hfill
        
        \caption{KDE of message topics for different IT communities}\label{fig:kde-it}
    \end{minipage}
    
\end{figure*}

The English-speaking communities exhibit a marked tendency towards insularity, as QAnon is a very closed community~\cite{gianniriotta,networkqanonuserdynamics}. Indeed, many communities, although primarily connected with QAnon themes, show a distinct emphasis on topics such as cryptocurrency, health, or governmental affairs, unified by an underling QAnon narrative. 
This phenomenon of thematic variations within a singular ideological framework is indicative of the QAnon community's cohesiveness. 
Indeed, prior work has observed an increasing association of QAnon with religiosity, alternative health, and wellness philosophies, as well as affective states that promote conspiratorial thinking~\cite{willaert2023computational,greer2024belief,tuters2022deep} -- trends also observed in the Italian-speaking communities.

\subsection{t-SNE for context analysis}

To provide a comprehensive visual representation of the topics discussed within our datasets, we represent all messages using t-Distributed Stochastic Neighbor Embedding (t-SNE)~\cite{JMLR:v9:vandermaaten08a}, a dimensionality reduction technique used for visualizing high-dimensional data through visual clustering. In this way, spatial proximity in the t-SNE map can suggest how topics fit into the larger conversation on conspiracies. We build the t-SNE visualization on topics identified by the CorEx algorithm. In particular, we developed two distinct models, one for Italian and one for English to analyze the entire corpus of messages. We opted to identify 50 topics to further our understanding of the context dynamics inside the clusters. By representing each message as a 50-dimensional vector corresponding to these topics, we can highlight the diverse contexts within each community. This is particularly important because Telegram chats often cover a broad range of topics rather than focusing on a single subject \cite{hoseini2023globalization}.
We obtain and $n\times m$ matrix where $n$ and $m$ are respectively the number of messages and the number of topics we wanted to detect. Each value $v_{i,j}$ represents the correlation between the $i\textsuperscript{th}$ message and the $j\textsuperscript{th}$ topic. 
We lower the dimensionality of our matrix using the tSNE and plot all messages in a two-dimensional space, coloring them according to the community of origin to show how clusters are closely related or share similar discussions.
Figure \ref{fig:tsne} presents the results on the English dataset. The varying distributions of the messages across communities highlight the differences in discussion in terms of quantity, focus, and framework, even among similar communities. This spatial arrangement underlines the nuanced interactions between these communities. For example, we can observe the proximity of the \texttt{QAnonCrypto} community to the \texttt{QAnon} and the \texttt{QAnonHealth} communities, suggesting that crypto topics tend to piggyback engage with QAnon-related discussions. Figure \ref{fig:kde-en} better presents the differences in distributions through Kernel Density Estimation (KDE) of the messages, where areas of higher density indicate a higher likelihood of encountering messages related to specific topics. For instance, in Figure \ref{fig:kde-en}a, the distribution of messages in chats of the \texttt{QAnon} community is notably widespread, suggesting correlations with many different topics, similarly to \texttt{QAnonCrypto} (Figure \ref{fig:kde-en}b) and \texttt{Warfare} (Figure \ref{fig:kde-en}c) communities. This suggests that some communities on Telegram tend to discuss a broad array of topics, they each enrich the discourse with their unique frameworks and worldviews.  In contrast, more specialized communities like \texttt{OldSchoolConsp} (Figure \ref{fig:kde-en}d) are localized to very specific areas. 
We conduct the same analysis for the Italian dataset. Due to space constraints, we highlight only some notable patterns. We observe distinct patterns between the \texttt{NewsSource} (Figure \ref{fig:kde-it}a) and \texttt{AltNews} (Figure \ref{fig:kde-it}b) communities, which both cover alternative news topics. However, \texttt{NewsSource} also includes legitimate news sources, resulting in messages that show dual density peaks, possibly indicating interdependence, whereas \texttt{AltNews} messages display a single density peak, reflecting a more homogeneous topic focus.

\section{Validation}
\begin{figure}[!t]
    \centering
    \includegraphics[width=\columnwidth]{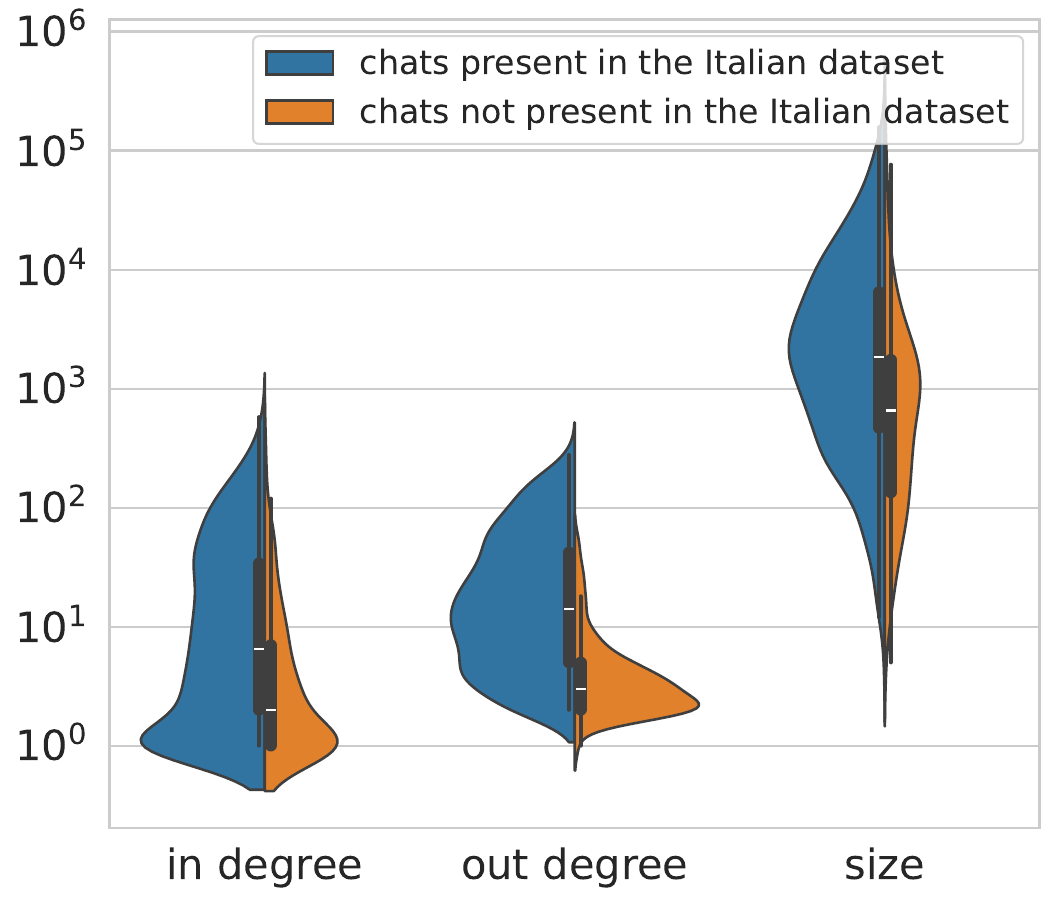}
    \caption{Difference in distribution of in-degree, out-degree, and size between the chats in the  validation dataset included in the initial Italian dataset and those that are not.}\label{fig:valid-param}
\end{figure}

Here, we show that the insights derived from our network analysis are not overly dependent on the initial seeds used to construct the dataset. This robustness check highlights the applicability of our methodology across different settings and its potential for broader research applications in the study of online discourse and information diffusion.
To assess the robustness of our findings, we aim to determine if starting from different seeds results in the same chat composition in our dataset. We focus on the Italian dataset and create a counterpart validation dataset using the snowballing process, this time starting from a distinct set of 28 seeds that were not among the original 43 Italian seeds used in the initial data collection. These new seeds are sourced from the Butac blacklist\footnote{\url{https://www.butac.it/the-black-list/}}, a list of Italian disinformation Telegram channels. The collected dataset includes 1,591 chats active from February 1, 2024 to March 20, 2024. We stopped the collection after two iterations of the process to maintain consistency with the original methodological framework. 
We then examine the overlap between the Italian datasets and the validation dataset to determine if the chats retrieved in the validation dataset match those in our original dataset. We find that $80\%$ of the chats in the validation dataset are also present in our original dataset, suggesting that our results would remain robust even with a different set of seeds. To further validate this finding, in Figure~\ref{fig:valid-param} we compare the size, in-degree, and out-degree distributions between chats included in our original dataset and those in the validation dataset that are not included in the original. The results indicate that the chats excluded from the original dataset have lower averages in size, in-degree, and out-degree, suggesting that the missing chats have less influence within the dataset.

\section{Discussions and limitations}


Leveraging the Telegram message forwarding mechanism has unveiled distinct trends and dynamics within conspiracy theory discussions across cultural contexts. In Italian-speaking communities, the diversity in handling conspiracy theories -- from challenging mainstream narratives with alternative information to sharing views from more traditional news sources -- enriches our understanding of the Italian conspiracy ecosystem on Telegram, a relatively uncharted territory in existing literature. 
The presence of news sources and alternative news outlets shows a dynamic interplay in the dissemination and legitimization of conspiracy theories, highlighting the intricate balance between mainstream credibility and the counter-narratives that thrive on Telegram. 
We also show trends of thematic diversity within a cohesive ideological framework, with conspiracy narrative ties to religiosity, alternative health, and conspiratorial thinking, trends similarly observed in literature for English-speaking groups~\cite{willaert2023computational,greer2024belief,tuters2022deep}.
The English-speaking communities span various topics like cryptocurrency, health, and governmental affairs, yet are tightly woven around the QAnon narrative~\cite{gianniriotta,networkqanonuserdynamics} with no presence of legitimate news sources, suggesting a significant echo chamber effect where misinformation may circulate more freely without the counterbalance of accredited information.
Our methodological robustness check suggests a relative independence from the choice of initial seeds used for the dataset construction. This implies that, despite starting from different seeds, we would likely have mapped out similar networks, suggesting that the communities identified through message forwarding -- encompassing both channels and groups -- tend to stay focused on conspiracy themes, thus remaining within this thematic bubble and fostering community homophily.
These observations align with previous studies indicating that channel communities engaged in forwarding tend to form echo chambers with varying structures~\cite{BovetGrindrod2022ukfarright}.

However, the diffusion of misinformation, a process inherently temporal and complex, cannot be fully captured through this static analysis alone. Future work should incorporate temporal network analyses to fully capture the actual journey of misinformation through the network or to uncover dynamic coordinated communities on Telegram~\cite{tardelli2023temporal}.
Despite this limitation, the insights and robustness check highlight the applicability of our methodology across different settings and its potential for broader research applications in the study of online discourse and information diffusion.

\section{Conclusions}

In this study, we analyzed online Italian and English conspiracy-related Telegram communities through the lens of message forwarding, aiming to uncover the dynamics of conspiracy theory discussions in different speaking contexts. Using snowball sampling, we collected two extensive datasets encompassing Telegram channels, groups, linked chats, and messages shared over a month in 2024. We built the Italian and English networks, revealing key communities, and characterize their narratives through topic modeling. We uncovered trends of thematic diversity within a cohesive ideological framework, linking conspiracy narratives to religiosity, alternative health, and conspiratorial thinking, and uncovered the interplay of news sources and alternative news outlets in disseminating and legitimizing conspiracy theories. Our analysis also shed light on the thematic relationships between communities and the role of forwarded messages in fostering content distribution and community homophily. Finally, we tested our methodology's robustness against variations in initial dataset seeds, showing the reliability of our insights and broader applicability.
This research contributes new perspectives on misinformation spread, paving the way for further exploration of conspiracy discourse, especially in the under-explored Italian context, and misinformation diffusion on Telegram.

\section*{Acknowledgments}
This work was partly supported by SoBigData.it which receives funding from European Union – NextGenerationEU – National Recovery and Resilience Plan (Piano Nazionale di Ripresa e Resilienza, PNRR) – Project: “SoBigData.it – Strengthening the Italian RI for Social Mining and Big Data Analytics” – Prot. IR0000013 – Avviso n. 3264 del 28/12/2021.; and by project SERICS (PE00000014) under the NRRP MUR program funded by the EU – NGEU.

\bibliographystyle{IEEEtran}
\bibliography{bibliography}

\end{document}